\documentstyle[twocolumn,aps]{revtex}

\begin{document}
\title{Quantum gates by coupled quantum dots
and measurement procedure in Si MOSFET}

\author{Tetsufumi Tanamoto}
\address{Corporate Research and Development Center,
Toshiba Corporation, Saiwai-ku, Kawasaki 210-8582, Japan}

\date{\today}
\maketitle

\begin{abstract}
We investigated the quantum gates of coupled quantum dots, 
theoretically, when charging effects can be observed. 
We have shown that the charged states in the qubits can be observed 
by the channel current of the MOSFET structure. 
\end{abstract}

\narrowtext

\section{Introduction}
Since Shor's factorization program was proposed, 
many studies have been carried out in order to 
realize the quantum computer
\cite{Ekert,Gershenfeld,Schon,Nakamura}. 
Nakamura succeeded in the control of the 
macroscopic quantum state as the solid devices of 
Josephson junctions\cite{Nakamura}.
Recently, we have proposed the quantum computer in 
the asymmetric coupled dot in Si nanocrystals\cite{tana9902}.
The advantage of using the coupled dots attached by the gate
electrode is that a quantum state can be controlled by the 
gate voltage in the solid circuits.  
Moreover, when  the coupled Si quantum dots are embedded in
the gate insulator of the MOSFET\cite{Tiwari,Guo}, 
the quantum state in the two coupled dots (qubits) is expected 
to be detected by the channel current.  
In the previous paper\cite{tana9902}, we investigated the case
where the sizes of Si nanocystals were small such that
the energy-levels in the quantum dots were discrete 
and the two-state system was constituted by the lowest 
energy-levels of the two quantum dots. 
Here, in this paper, we consider the case where the sizes
of the quantum dots are larger (of the order of a few tens of nm) 
and the energy-levels of the individual quantum dots 
are almost continuous\cite{Tsu2}.
We also assume that the charging energy of each tunneling
junction is large enough that the Coulomb blockade effects
can be seen.
This was realized in the quantum gate of the Josephson junction
by Shnirman\cite{Schon} and Averin\cite{Averin}. 
The superconducting state has the advantage from the 
viewpoint of the decoherence that 
the qubits are connected by the coherent circuits.
On the other hand, we use the normal state of the 
capacitively coupled quantum dots, because 
there exist some predictions that the decoherence time 
in the normal quantum dot array in semiconductors 
is not expected to be so short\cite{Zanardi}. 
First, in this paper, we show that the capacitively coupled dots 
can be reduced to a two-state system of interacting two qubits.
Next, we show the detecting mechanism of the 
quantum state by the channel current based on 
the conventional MOSFET model. 
Asymmetry of the coupled dots is not assumed and we set $e$=1.

\section{Capacitively coupled quantum dots as two qubits}
The configuration of the quantum dots and capacitances 
are shown in Fig.\ref{fig1}. 
We assume that the coupling between qubits 
is weaker than that in a qubit(The full 
formation of the electrostatic energy is shown 
in the Appendix), and we take
\begin{equation}
C_1\!=\!C_2\!=\!\!\sqrt{2}C_3\!=\!\!\sqrt{2}C_4 \ll 
C_5(\!=\!C_8), C_6(\!=\!C_9), C_7(\!=\!C_{10}).
\end{equation}
By this assumption, we can expand the charging energy 
as a function of $n_a\equiv N_{\rm A}-N_{\rm B}$ 
and $n_b\equiv N_{\rm C}-N_{\rm D}$ 
($N=N_{\rm A}+N_{\rm B}=N_{\rm C}+N_{\rm D}$),
by a small $C_3$, and obtain the electrostatic energy:
\begin{eqnarray}
U(n_a, n_b) \!&=&\!E_c
\left[n_a \!+\! \frac{\eta}{2E_c} n_b \!+\!
\frac{C_7\!-\!C_5}{C_5\!+\!C_7} N
\!-\!\frac{2C_5 C_7}{C_5\!+\!C_7} V_a \right]^2 
\nonumber \\ \!&+&\!
E_c\left[n_b\!+\!\frac{C_7\!-\!C_5}{C_5\!+\!C_7} N
\!-\!\frac{2C_5 C_7}{C_5\!+\!C_7} V_b \right]^2
\label{eqn:U_1}
\end{eqnarray}
where $E_c\equiv (C_5+C_7)/(8(C_5C_6+C_6C_7+C_7C_5))$ and  
\begin{equation}
\eta \!=\! \frac{\sqrt{2}(C_5^2\!+\!C_7^2\!- \!\sqrt{2}C_5C_7)}
{4(C_5 C_6\!+\!C_6 C_7\!+\!C_7 C_5)^2} C_3
\end{equation}
If we consider the gate voltage region where the 
electrostatic energy of $n_j=0(j=a,b)$ state crosses that of 
$n_j=1$ state described by Shnirman\cite{Schon} 
and Averin\cite{Averin}, 
the electrostatic energy and the tunneling amplitude, $\Omega_j$, 
constitute the Hamiltonian of the two-state system as
\begin{equation}
H=\sum_{j=a,b}(\epsilon_j \sigma_{zj}
+\Omega_j \sigma_{xj}) -(\eta/4) \sigma_{za}\sigma_{zb}
\end{equation}
where $\epsilon_a=E_c(1/2+[(C_7-C_5)N-2C_5C_7V_a]/(C_5+C_7))$ 
and $\epsilon_b=E_c(1/2+[(C_7-C_5)N-2C_5C_7V_b]/(C_5+C_7))$. 
$\sigma_x$ and $\sigma_z$ are Pauli matrices.  
When the capacitances are approximated as 
$C_i = 2 \pi \epsilon_{\rm ox}r^2/
(d_i + (\epsilon_{\rm ox}/\epsilon_{\rm Si}) r)$
where $\epsilon_{\rm ox}=4$, $\epsilon_{\rm Si}=12$, 
$d_i$ is the size of the capacitance  and
$r$ is the radius of each quantum dot, 
then $V_a,V_b$ is of the order of tens of meV.

The mechanism of the controlled NOT operation is similar
to that of the Josephson junction\cite{Averin}. 
Whether $n_b=0$ or $n_b=1$,  the level-crossing gate voltage, 
$V_a$, shifts(see Eq.(\ref{eqn:U_1}) 
and the controlled NOT operation is realized.

The dynamical motion of the excess charge in the 
two-state system can be easily considered by solving 
the time-dependent Schr\"{o}dinger equation\cite{Tsukada}, 
and the excess charge shows the oscillating behavior depending 
on the energy of the two states. 
The time period of the oscillation, $\tau_\delta$, 
can be simply approximated as 
$\tau_\delta \sim \hbar/\sqrt{\Omega^2 + (\epsilon_a -\epsilon_b)/4}$. 

\section{Dissipation by environmental phonons}
The polarized charged state of a qubit (coupled dots) 
behaves as the dipole moment under the electric field 
generated by the gate electrode. 
By the coupling of the dipole with the electric field, 
the two-state system can be described by the Bloch equation
and the quantum calculation is realized 
similar to the NMR quantum computer\cite{Gershenfeld}. 
It is well known that the one of the attractive characteristics 
 of the NMR quantum computer is its long decoherence  time. 
Here, we comment on the decoherence time of the semiconductor dot array. 
The decoherence time in semiconductor 
dot array has been considered to be short. This is 
the reason why Shnirman\cite{Schon}
and Averin\cite{Averin} used the Josephson effects in 
quantum gates. 
The decoherence in this case is considered mainly
to originate from the phonon environments, 
where the interaction between the 
two-state system and the phonon bath is largely given by 
a deformation potential\cite{Garcia}. 
The estimated decoherence time 
is not so short and of the order of $10^{-7}$ sec
from the analysis based on the model 
of Leggett\cite{Leggett}. 
This relatively long decoherence time will be 
related to the 'phonon bottleneck' derived by 
Zanardi\cite{Zanardi}. 
We will have to include the effects of the higher 
excited energy-levels and temperature for more detailed 
estimation. 

\section{Measurement mechanism in MOSFET}
The qubit which changes the charge distribution 
can be detected by the MOSFET structure. 
MOSFET structure\cite{tana9902} is considered to 
be a more efficient 
detecting devise in semiconductor quantum dots 
compared to the SET structure\cite{Shnirman}.
This is due to the change in the charge distribution 
in the qubit being detected by the capacitance effects 
similar to those 
of the quantum point contact\cite{Gurvitz,Hackenbroich}.
In this section, we show the detailed detecting mechanism of 
the MOSFET based on the long-channel MOSFET model 
in the case of two qubits.
The qubit system in the MOSFET proposed here can be seen as 
series of single coupled-dot MOSFETs (Fig.\ref{fig3}). 
When bias, $V_D$, is applied between the source and drain, 
the depletion region expands from the source and drain 
such that the width of the depletion region increases 
toward the drain. 
Thus, the channel current differs depending on the positions 
of the qubits which change the charge distribution. 
The channel current between the $i$-th qubit and $(i-1)$-th qubit  
is given for a small $V_D$ region as\cite{Sze},
\begin{equation}
I_D^{(i)} 
\sim \beta_0 \!
\left(
[V_{G}^{(i)}\!-\!V_{\rm th}^{(i)}](V_i\!-\!V_{i-1})\!-\!
\frac{1}{2} \alpha 
(V_i^2\!-\!V_{i-1}^2)
\right),
\end{equation}
where $\beta_0\equiv Z\mu_0 C_0/L_{i}$
($Z$ is the channel width, $\mu_0$ is the mobility, $L_{i}$
is the channel length of $i$-th qubit where we set 
$L_1\!=\!L_2\!=\!\cdot \cdot \cdot \!=\!L_{\rm N}$, and 
$C_0$ is the capacitance of the SiO${}_2$) and  
$\alpha \equiv1\!+\!\frac{1}{4\varphi_{\rm B}}
\frac{Q_{B}}{C_0}$ where 
$Q_{\rm B}$ is the charge within the surface depletion 
region.  
$V_{\rm G}$ is the gate voltage, and 
the threshold voltage,  $V_{\rm th}^{(i)}$, is 
given by $ V_{\rm th}^{(i)}\!=\!V_{\rm th}
\!+\!\Delta V_{\rm th}^{(i)}$ 
where $V_{\rm th}\equiv V_{\rm FB}+2\varphi_{\rm B}+
\frac{Q_{B}}{C_0}$ ($V_{\rm FB}$ is a flat band voltage, 
$\varphi_{\rm B}$ is the potential difference 
between the Fermi level and the intrinsic Fermi 
level of substrate), and the shift 
by the change of the charge distribution, $\Delta V_{\rm th}^{(i)}$ 
in the $i$-th qubit.
Then, we have the following conditions:
\begin{equation}
V_N=V_{DS} \ \ \ {\rm and} \ \ \  
I_D^{(1)}=I_D^{(2)}= \cdot  \cdot  \cdot =I_D^{(N)}
\end{equation}
In the case of two qubits, 
with $V_{Gi}=V_{G}^{(i)}-V_{\rm th}^{(i)}$ 
($V_{Gi}\gg V_D$ is assumed),
\begin{equation}
I_D =
\frac{\beta_0}{(V_{G1}+V_{G2})}
(V_{G1}V_{G2} V_D-\frac{\alpha V_{G1}}{2}V_D^2)
\end{equation}
Thus, whether 
($V_{G1}=V_g-\Delta V_{\rm th}$ and $V_{G2}=V_g$ ) or 
($V_{G1}=V_g$ and $V_{G2}=V_g-\Delta V_{\rm th}$ ),  
the difference of the corresponding currents, 
$\Delta I_D^{(12)}$ is given as
\begin{equation}
\Delta I_D^{(12)} \approx \frac{\beta_0 \alpha}
{2(2V_g -\Delta V_{\rm th})}
\Delta V_{\rm th} V_D^2.
\end{equation}
This difference can be observed in the
nonlinear $I_D$-$V_D$ region and, 
in the pure linear region where the 
terms which include $\alpha$ disappear, 
the changed qubits cannot be  distinguished. 

\section{Conclusions} \label{sec:conclusion}
We have investigated the quantum gates of the capacitively coupled 
quantum dot array in the MOSFET structure, and 
have derived the two-state 
Hamiltonian by the capacitances of the quantum dots. 
We also have shown the detecting mechanism of the MOSFET 
structure by analyzing the channel current based on the 
conventional MOSFET model. 

\acknowledgments
The author is grateful to N. Gemma, S. Fujita, K. Ichimura, 
 and J. Koga for fruitful discussions.

\appendix
\section{The electrostatic energy of the two qubits}
In this section we show the electrostatic energy of the 
two qubits in terms of the capacitances of dots and the 
gate electrodes (Fig. \ref{fig1})\cite{Matveev,Waugh}. 
We assume $C_1\!=\!C_2\!=\!\sqrt{2}C_3\!=\!\sqrt{2}C_4$ , 
$C_8\!=\!C_5$, $C_9\!=\!C_6$ and $C_{10}\!=\!C_7$.
The total electrostatic energy of the two qubits is given by
\begin{equation}
U=\sum_{i=1,...,10}\frac{q_i^2}{2C_i} -q_7 V_a -q_{10} V_b, 
\label{app1}
\end{equation}
where $q_i$ shows the charge of the $i$-th capacitance 
and we have the relation between $q_i (i=1,2...10)$ and 
the total charge of the fore dots, $N_{\rm A}$, ...,$N_{\rm D}$ 
as
\begin{eqnarray}
-N_{\rm A}&\!=\!& q_1\!+\!q_3\!+\!q_5\!+\!q_6, \ \ 
-N_{\rm B}\!=\! q_2\!+\!q_4\!-\!q_5\!+\!q_7, \nonumber \\
-N_{\rm C}&\!=\!&\!-\!q_1\!-\!q_4\!+\!q_8\!+\!q_9, \ \ 
-N_{\rm D}\!=\! \!-\!q_2\!-\!q_3\!-\!q_9\!+\!q_{10}. \nonumber 
\end{eqnarray}
By minimizing the energy Eq.(\ref{app1}) at the fixed 
values of $V_a$, $V_b$ and $N_{\rm A} ...N_{\rm D}$, 
we have
\small
\begin{eqnarray}
\lefteqn{U(n_a, n_b) \!=\! 
\frac{1}{16}(\frac{1}{C_a}\!+\!\frac{1}{C_b}) n_a^2
\!+\!\frac{1}{4}\! \left\{\! \frac{C_7\!-\!C_5}
{\!C_A C_B\!-\!(C_3\!+\!C_6\!)^2}N
\!-\!\frac{C_7C_5}{\!C_A C_B\!-\!(C_3\!+\!C_6\!)^2}V^+\!
\!+\!\frac{C_6\!-\!C_3\!-\!C_C}
{\!C_C C_D\!-\!(C_6\!-\!C_3\!)^2}V^- 
\right\}n_a  }\nonumber \\
&+& \frac{1}{16}(\frac{1}{C_a}\!+\!\frac{1}{C_b})  n_b^2
+\frac{1}{4} \left\{ \frac{C_7\!-\!C_5}{C_A C_B\!-\!(C_3\!+\!C_6)^2}N
\!-\!\frac{C_7C_5}{C_A C_B\!-\!(C_3\!+\!C_6)^2}V^+\!
-\!\frac{C_6\!-\!C_3\!-\!C_C}{C_C C_D\!-\!(C_6\!-\!C_3)^2}V^- 
\right\}n_b  \nonumber \\
\!&+&\!\frac{1}{8}(\frac{1}{C_a}\!-\!\frac{1}{C_b}) n_a n_b
\!+\!\frac{1}{4C_A}N^2  \!+\!
\frac{C_A}{4(\!C_A C_B\!\!-\!\!(\!C_3\!\!+\!\!C_6\!)^2)} \!
\left[\frac{C_3\!\!+\!\!C_6\!\!+\!\!C_A}{C_A} N \!\!
+\!\! C_7 V^+\right]^2 \!\!\!+\!\! 
\frac{C_C}{4(\!C_C C_D\!\!-\!\!(\!C_6\!\!-\!\!C_3\!)^2)}
( C_7 V^+)^2 \nonumber
\end{eqnarray}
\normalsize
where 
\begin{eqnarray}
C_A \!&=\!&  C_3 \!+\!C_5\!+\!C_6, \ \ C_B 
\!=\!C_3\!+\!C_6\!+\!C_7, \nonumber \\
C_C \!&=\!& \varrho C_3\!+\!C_5\!+\!C_6, \ \
C_D \!=\! \varrho C_3\!+\!C_6\!+\!C_7,\nonumber
\end{eqnarray}
and $V^{\pm}=V_a \pm V_b$, 
$1/C_a=(C_5 +C_7)/( (C_5+C_7)(C_3+C_6)+C_5 C_7)$,
$1/C_b=[C_5 +C_7+2(\varrho+1) C_3]/
[(C_5 +C_6+\eta C_3)(C_6+C_7+\varrho C_3)-(C_6-C_3)^2]$ 
and $\varrho=2\sqrt{2}+1$.


\begin{figure}
\caption{Two coupled qubits by the quantum coupled dots.}
\label{fig1}
\end{figure}

\begin{figure}
\caption{
An example of the $N$ coupled dot system of quantum computing. 
Dots are coupled in the longitudinal direction. 
The electron transfer in the lateral direction is assumed to be 
negligible. 
The FET channel structure enables the detection of the small 
signal of the charge distribution in coupled quantum dots.}
\label{fig3}
\end{figure}

\end{document}